\documentclass[12pt,preprint]{aastex}
\usepackage{apjfonts}

\def\wig#1{\mathrel{\hbox{\hbox to 0pt{%
          \lower.5ex\hbox{$\sim$}\hss}\raise.4ex\hbox{$#1$}}}}

\shorttitle{Planetary Radii}
\shortauthors{Fortney, Marley, \& Barnes}

\newcommand{\mj}{$M_{\mathrm{J}}$}
\newcommand{\rj}{$R_{\mathrm{J}}$}
\newcommand{\me}{$M_{\oplus}$}
\newcommand{\re}{$R_{\oplus}$}
\newcommand{\T}{TrES-1}
\newcommand{\hd}{HD 209458b} 
\newcommand{\hh}{HD 149026b}

\newcommand{\ha}{HAT-P-1b}

\newcommand{\ct}{\citet}

\slugcomment{\bf}
\slugcomment{ApJ in print, April 17, 2007}

\begin{document}

\title{Planetary Radii across Five Orders of Magnitude in Mass and Stellar Insolation:  Application to Transits}

\author{J. J. Fortney\altaffilmark{1}$^,$\altaffilmark{2}$^,$\altaffilmark{3}, M. S. Marley\altaffilmark{1}, J. W. Barnes\altaffilmark{4}}

\altaffiltext{1}{Space Science and Astrobiology Division, NASA Ames Research Center, MS 245-3, Moffett Field, CA 94035; jfortney@arc.nasa.gov, mark.s.marley@nasa.gov}
\altaffiltext{2}{Spitzer Fellow}
\altaffiltext{3}{Carl Sagan Center, SETI Institute, 515 North Whisman Road, Mountain View, CA 94043}
\altaffiltext{4}{Lunar and Planetary Laboratory and Department of Planetary Sciences, University of Arizona, Tucson, AZ, 85721; jbarnes@c3po.lpl.arizona.edu}

\begin{abstract}

To aid in the physical interpretation of planetary radii constrained through observations of transiting planets, or eventually direct detections, we compute model radii of pure hydrogen-helium, water, rock, and iron planets, along with various mixtures.  Masses ranging from 0.01 Earth masses to 10 Jupiter masses at orbital distances of 0.02 to 10 AU are considered. For hydrogen-helium rich planets, our models are the first to couple planetary evolution to stellar irradiation over a wide range of orbital separations (0.02 to 10 AU) through a non-gray radiative-convective equilibrium atmosphere model.  Stellar irradiation retards the contraction of giant planets, but its effect is not a simple function of the irradiation level: a planet at 1 AU contracts as slowly as a planet at 0.1 AU.   We confirm the assertion of Guillot that very old giant planets under modest stellar irradiation (like that received by Jupiter and Saturn) develop isothermal atmospheric radiative zones once the planet's intrinsic flux drops to a small fraction of the incident flux.  For hydrogen-helium planets, we consider cores up to 90\% of the total planet mass, comparable to those of Uranus and Neptune.  If ``hot Neptunes'' have maintained their original masses and are not remnants of more massive planets, radii of $\sim$0.30-0.45 \rj~are expected. Water planets are $\sim40-50$\% larger than rocky planets, independent of mass.  Finally, we provide tables of planetary radii at various ages and compositions, and for ice-rock-iron planets we fit our results to analytic functions, which will allow for quick composition estimates, given masses and radii, or mass estimates, given only planetary radii. These results will assist in the interpretation of observations for both the current transiting planet surveys as well as upcoming space missions, including COROT and Kepler.

\end{abstract}


%\newpage

\section{Introduction}

We are still in the early days of a revolution in the field of planetary sciences that was triggered by the discovery of planets around other stars.  Exoplanets now number over 200, with masses as small as $\sim$5-7 \me~\citep{Rivera05,Beaulieu06}.  Comparative planetology, which once included only our solar system's planets and moons, now includes sub-Neptune to super-Jupiter-mass planets in other solar systems.

Currently the most important class of exoplanets are those that transit the disk of their parent stars, allowing for a determination of planetary radii.  The 14 confirmed transiting planets observed to date are all more massive than Saturn, have orbital periods of only a few days, and orbit stars bright enough such that radial velocities can be determined, allowing for a calculation of planetary masses and bulk densities \citep[see][]{Charb06}.  A planetary mass and radius allows us a window into planetary composition \citep{Guillot05}.  The 14 transiting planets are all gas giants (See \citealt{Guillot96, Guillot02, Bodenheimer03, Burrows03, Baraffe05, Laughlin05, Arras06} for calculations of structure and contraction of many planets) although one planet, \hh, appears to be $\sim$2/3 heavy elements by mass \citep{Sato05, Fortney06, Ikoma06}.  Understanding how the transiting planet mass-radius relations change as a function of orbital distance, stellar mass, stellar metallicity, or UV flux, will provide insight into the fundamentals of planetary formation, migration, and evolution.  Tentatively, some work in this general direction is beginning \citep{Guillot06}.

The transit method of planet detection is biased towards finding planets that orbit relatively close to their parent stars.  This means that radial velocity followup will be possible for some planets as the stellar ``wobble'' signal is larger for shorter period orbits.  However, for transiting planets that are low-mass, or that orbit very distant stars, stellar radial velocity measurements may not be possible.  For planets at larger orbital distances, radial velocity observations may take years.  Therefore, for the foreseeable future a measurement of planetary radii will be our only window into the structure of these planets.  Estimates of masses will still be important, however, for useful constraints on planet formation theories.  This will have to involve some degree of presumption regarding the composition of planets.  Orbital distances may give some clues as to a likely composition, but our experience over the past decade with Pegasi planets (or ``hot Jupiters'') has shown us the danger of assuming certain types of planets cannot exist at unexpected orbital distances.  Here we compute planetary radii as function of mass, composition, and stellar irradiation to help in these future planetary mass and composition estimates.

In \S2 we give a brief overview of the COROT and Kepler missions.  In \S3 we describe the scope of the calculations and our aims in this study.  Later, \S4 describes our methods while \S5 and \S6 give our results for ice-rock-iron planets and gaseous planets, respectively.  Finally, \S7 is our conclusions and suggestions for future work. 

\section{Upcoming Space Missions}
The French/European COROT mission, set to launch in December 2006, and American Kepler mission, set to launch in November 2008, will revolutionize the study of exoplanets.  COROT will monitor 12,000 stars in each of 5 different fields, each for 150 continuous days \citep{Borde03}.  Planets as small as 2 \re~should be detectable around solar-type stars \citep{Moutou06}.  The mission lifetime is expected to be at least 2.5 years.  The Kepler mission will continuously monitor one patch of sky, monitoring over 100,000 main sequence stars \citep{Basri05}.  The expected mission lifetime is at least 4 years.  Detection of sub-Earth size planets is the mission's goal, with detection of planets with radii as small at 1 Mercury radius possible around M stars.   With these missions, perhaps hundreds of planets will be discovered with masses ranging from sub-Mercury to many times that of Jupiter.  Of course while planets close to their parent stars will preferentially be found, due to their shorter orbital periods and greater likelihood to transit, planetary transits will be detected at all orbital separations.  In general, the detection of three successive transit will be necessary for a confirmed detection, which will limit confirmed planetary-radius objects to $\sim$1.5 AU.  It is important to remember that COROT and Kepler will not determine planetary masses, only radii, so until followup radial velocity or astrometry work is done, planetary radii will be our only window into the composition of these new planets.

\section{Focus \& Scope of Calculations}
In the next few years technology will allow the detection of transiting planets that range from many Jupiter masses (1 \mj = 317.89 \me), to perhaps as small as 0.01 \me.  Interpreting these observations will require planetary radii to be understood over more than five orders of magnitude in mass.  Discoveries to date have been surprising, including transiting planets that are larger than standard models can explain (such as \hd~and \ha), Saturn-mass planets that have four Neptune-masses worth of heavy elements (such as \hh), and amongst those that do not transit, Neptune-mass planets that are quite hot and perhaps not ice-rich (such as HD 68930b), and $\sim$5-10 \me~planets that some are calling ``super-Earths'' (such as Gliese 876d).  Therefore, we think it is useful to take as broad a view as possible.  We make few assumptions regarding composition, and we calculate radii of pure hydrogen-helium planets as small as 10 \me, water and rock dominated planets up to 1000 \me ($\sim$3 \mj), along with many compositions in between.  Since we have the theoretical tools at our disposal, excluding some compositions a priori is unnecessary at this point.

We use state-of-the-art equations of state (EOS) for iron \citep[Sesame 2140,][]{Sesame}, olivine (for generic ``rock'') \citep[ANEOS,][]{ANEOS}, water \citep[ANEOS,][]{ANEOS}, helium \citep{SCVH}, and hydrogen \citep{SCVH}.  Although detailed models for the thermal evolution of super terrestrial planets have been published \citep{Valencia06,Ehrenreich06}, issues discussed in such detailed models, such as lithospheric thickness and interior temperature structure have only a small effect on planetary radii \citep[temperature effects may reach $\sim$4\% in radius for the highly irradiated ``Super Mercuries" of][]{Valencia06} and are unobservable in the forseeable future.  Here we are interested in the most readily observed quantity: planetary radii.  Even for precisely determined light curves, planetary radii deteminations are somewhat imprecise owing to uncertainties in parent star radii, which can often reach 10\%.

For hydrogen/helium planets, irradiation from a parent star is a significant energy source that strongly affects evolution and contraction, and must be accounted for.  We accurately incorporate stellar heating into our evolution (contraction) calculations through self-consistent non-gray model atmospheres under irradiation from 0.02 to 10 AU.  This is the first investigation of the evolution and contraction of hydrogen/helium planets under irradiation over such a wide range of orbital separations.

Other authors have previously computed radii for some subset of the compositions here.  \citet{Zapolsky69} is perhaps the best known, in which they calculated the radius of zero-temperature spheres of H, He, C, Fe, and Mg for masses from 0.3 \me~to 10 M$_\odot$.  Their Thomas-Fermi-Dirac EOSs are accurate at pressures where some degree of ionization takes place but  less so for the lower pressures of our solar system's terrestrial planets.  Similary, \citet{Stevenson82b} calculated radii from 1 to 1000 \me~for cold and warm H, H/He, ice, and rock planets.  \citet{Valencia06} have recently investigated the radii of Earth-like and Mercury-like planets up to 10 \me.  \citet{Saumon96} have calculated planetary radii of giant planets with and without cores.  \citet{Guillot96} also investigated the radii of 51 Peg b-like planets, with compositions ranging from H/He to rock.  More recently \citet{Bodenheimer03} have calculated radii of H/He planets with and without cores, at various orbital separations, but these models lack realistic atmospheric boundary conditions and equations of state for core materials.  Our results for water, rock, and iron bodies do not differ substantially from these other works, but we compute accurate radii over a wider range of masses, and also include mixed compositions.  However, our calculations for the radii of H/He rich planets are an important improvement over other works, as we explicitly and accurately include irradiation from a parent star.  In addition, we include the effects on planetary radii across a wide range of core masses.

\section{Methods}
\subsection{Equations of State} \label{eos}
\begin{figure}
\plotone{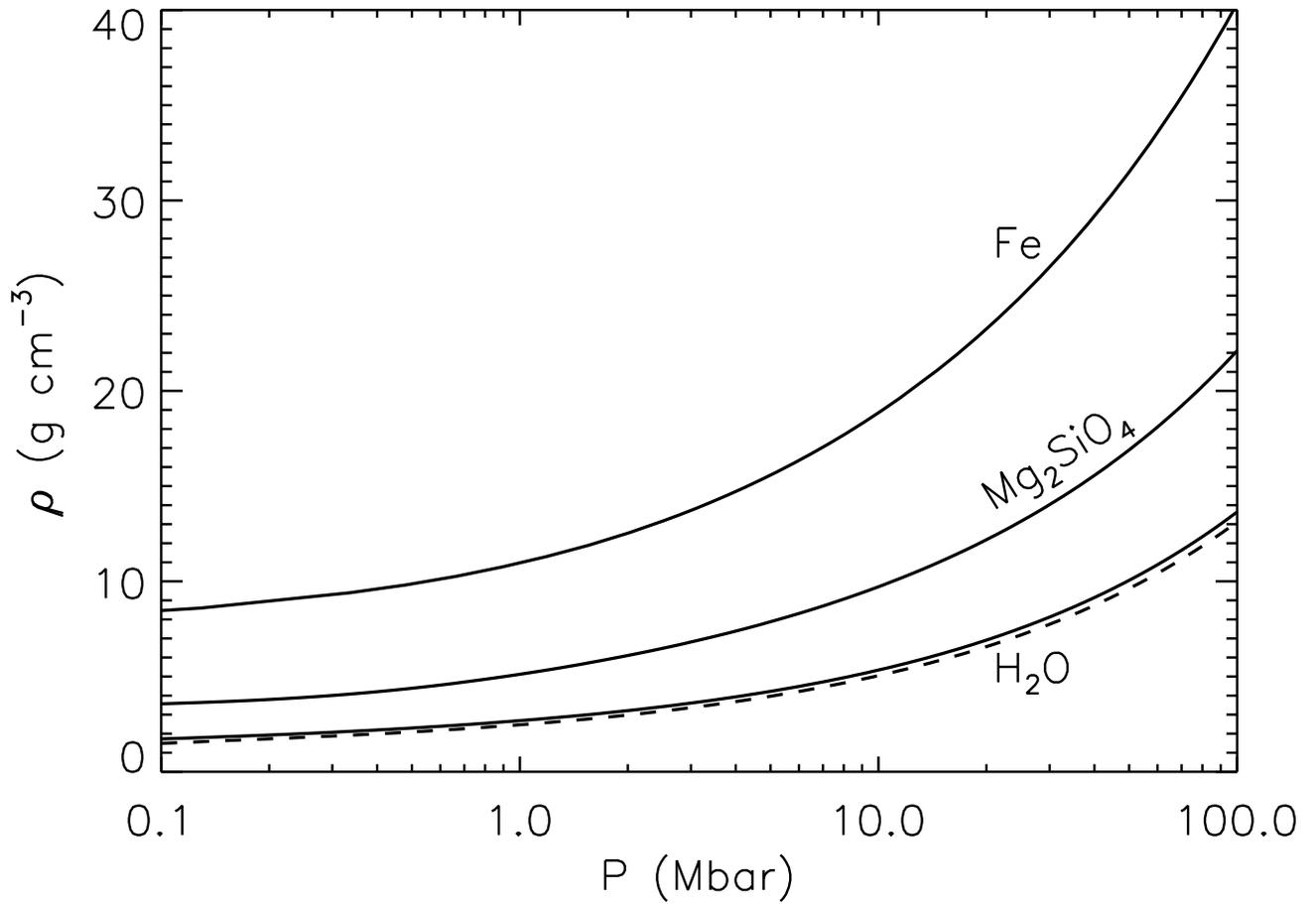}
\caption{Zero-temperature pressure-density relations for iron (Fe), rock (Mg$_2$SiO$_4$), and water ice (H$_2$0).  For ice, the 
dashed curve shows our EOS with the thermal correction described in \S\ref{eos}.
\label{fig:eos}}
\end{figure}

First we will look in a bit more detail into our equations of state before elaborating on the construction of the planet models.  The total pressure, ($P$), can be expressed as: 
\begin{equation}
P=P_0+P_{\mathrm{T}}
\label{p}
\end{equation}
where $P_0$ is the pressure at zero-temperature and $P_T$ is the thermal pressure.  For the degenerate interiors of massive planets, thermal effects are quite small for high atomic number species.  For instance, zero-temperature equations of state for rock are thought to be accurate to within $\sim$1-2\% for use in Uranus and Neptune interior models \citep{Hubbard80b, Hubbard84b}, so we make no thermal corrections for rock and iron.  For water, thermal pressure can be important at the $\sim$10\% level.  \citet{Hubbard80b} find a relation for the $P_{\rm T}$ of water that is suitable for the interior pressure-temperature (\emph{P--T}) profiles of Uranus and Neptune.  This relation,
\begin{equation}
\label{ice}
P_{\mathrm{T}}=3.59 \times 10^{-5} \rho T,
\end{equation}
where $P_{\rm T}$ is in Mbar, $\rho$ is in g cm$^{-3}$, and $T$ in K, is relevant for ``typical planetary interior conditions.''  For $\rho \sim 4$ g cm$^{-3}$ and $T \sim 5000$ K, $P_{\rm T} \sim 0.7$ Mbar, or 10\% of the total pressure, $P$.  For planets composed of any fraction of water, we assume that interior temperatures within the ice follow the Uranus/Neptune adiabat of \citet{Guillot05}, which reaches $\sim$550 K at 1 kar and 4000 K at 1 Mbar, and add the thermal pressure correction of Equation (\ref{ice}) at every $P_0$.  This is our EOS of ``warm ice.''  For the time being we will ignore the additional complexity of including the EOS of methane and ammonia, solar system ``ices'' that are not as abundant as water.  These molecules condense at colder temperatures and it is not clear how abundant these species may be in transiting planets, which are preferentially found relatively close to their parent stars.  In \mbox{Figure~\ref{fig:eos}} we show our equations of state for cold ice, warm ice, rock, and iron.  For hydrogen/helium envelopes, we compute internal adiabats with a helium mass fraction $Y=0.28$ and do not include heavy elements.  A description of this detailed H/He EOS can be found in \citet{SCVH}.

\subsection{Planetary Structure and Evolution}
The structure of spherically symmetric planets in hydrostatic equilibrium follow the relations set out below.  Equations (\ref{dm1}) and (\ref{dm2}) define mass conservation and hydrostatic equilibrium, respectively.  Equation (\ref{dm3}) defines energy conservation, which is employed in our evolution calculations of planets with a H/He envelope.
\begin{equation}
\label{dm1}
\frac{\partial r}{\partial m} = \frac{1}{4 \pi r^2\rho}
\end{equation}
\begin{equation}
\label{dm2}
\frac{\partial P}{\partial m} = \frac{-Gm}{4 \pi r^4}
\end{equation}
\begin{equation}
\label{dm3}
\frac{\partial L}{\partial m} = -T\frac{\partial S}{\partial t}.
\end{equation}
Here $r$ is the radius of a mass shell, $m$ mass of a given shell, $\rho$ the local mass density, $P$ the pressure, $G$ the gravitational constant, $L$ the planet's intrinsic luminosity, $T$ the temperature, $S$ the specific entropy, and $t$ the time.

For planets composed only of water, rock, or iron, we do not utilize Equation (\ref{dm3}), as we assume a constant radius with age.  Given the small thermal component of the pressure for these materials, and the expected uncertainty in radius measurements, this assumption is valid.  For planets where hydrogen and helium make up an appreciable mass fraction, following the thermal evolution and contraction of these planets is essential.  We note that in these models we do not include additional interior energy sources such as tidal dissapation.  This may be important for the hot Jupiters.  We also neglect helium phase separation, which will add $\sim$1000 km in radius to cold giant planets at Gyr ages \citep{FH04}.  Recall that for the planet with the most precisely determined radius \hd, the 1$\sigma$ radius uncertainty is still 1.9\% ($\sim$1800 km), due to uncertainties in the stellar parameters \citep{Knutson06}.

Our evolution code for the calculation of the cooling and contraction of adiabatic giant planets is well-tested.  It has been used to produce evolutionary models of Jupiter and Saturn \citep{FH03}, cool extrasolar giant planets \citep{FH04,Marley07}, hot Jupiters \citep{Fortney06}, and it is described in detail in \citet{FH03} and \ct{Fortney04}.  For all of these planets, it is the radiative atmosphere that serves as the bottleneck for cooling above the adiabatic H/He envelope.  This is accounted for with our fully non-gray, self-consistent model atmosphere grids.  The importance of using detailed atmosphere models for evolutionary calculations of hot Jupiters is discussed in \citet{Baraffe03} and \citet{Marley06}.

Below the H/He envelope we assume that heavy elements are found within a distinct core.  To model this core we use the EOS of a 50/50 by mass ice/rock mix using the ANEOS zero-temperature water and olivine EOSs \citep{ANEOS}.  The compositions of the cores of Jupiter and Saturn are not known, and given that we model planets that likely formed at a variety of orbital distances, in which different ratios of ice/rock could be accumulated, this simple choice is a reasonable one.  We ignore the heat content of the core on the thermal evolution of the planets.  This is often done for evolutionary models of Jupiter and Saturn \citep{Hubbard77,Saumon92,FH03}, as the error involved is small compared to other unknowns.  Please see \citet{Fortney06} for additional discussion on this point.

We also neglect the ``transit radius'' effect: The apparent radius of a transiting planet is the radius where the slant optical depth through the planet's atmosphere reaches unity.  The corresponding atmospheric pressure can vary across many orders of magnitude, depending on the wavelength \citep{Hubbard01,Fortney03}.  \citet{Burrows03} and \citet{Baraffe03} have estimated this effect to be $\sim$10\% and 5\% respectively, for \hd, compared to some reference radius, such as the radiative-convective boundary or the 1 bar level.  The \emph{HST} light curve of \citet{Brownetal01} and \citet{Charb02} was obtained in a narrow wavelength band that overlaps the strong sodium D line absorptions at 589 nm.  Across a broad visible wavelength band, for most planets, the transit radius effect would likely be only a few percent.  For the models presented here the radii correspond to a pressure of 1 bar.  Based on models from \citet{Burrows07}, an extension of $\sim$5-6 atmosphere scale heights from the 1 bar level is needed to reach the optical transit radius.

\subsection{Atmosphere Grids for Hydrogen-Helium Planets}
Giant planets have been shown to be fully convective, or nearly so, beneath their thin radiative atmospheres \citep[for a review, see][]{Hubbard02}.  The convection is thought to be quite efficient, and hence it is the radiative planetary atmosphere that serves as the bottleneck for escaping radiation and controls the cooling and contraction of the interior \citep{Hubbard77}.  As giant planet atmospheres have a number of atomic and molecular absorbers, including water, ammonia, methane, sodium, and potassium, these atmospheres are far from blackbodies \citep{Burrows97,Marley99,Sudar00}.  A model atmosphere grid, which serves as the upper boundary condition in these evolution calculations, relates the specific entropy ($S$) of the planet's internal adiabat and atmospheric surface gravity ($g$) to the planet's effective temperature ($T_{\rm{eff}}$).  While $S$ and $g$ are calculated from the planetary structure, $T_{\rm{eff}}$ can only be accurately determined from a non-gray planetary atmosphere code.

To compute the boundary condition for the evolution we compute self-consistent radiative-convective equilibrium atmospheric structure models on a large grid of gravities, intrinsic effective temperatures, and incident fluxes.  Each model computes an atmospheric temperature structure, accounting for deposition and re-radiation of incident starlight and convective transport and emission of internal thermal energy.  The model, based on one developed for Titan \citep{Mckay89}, was originally applied to the study of giant planet thermal structure by \citet{MM99} and has also been employed to study the atmospheres of the hot Jupiters \citep{Marley98,Fortney05, Fortney06} as well as brown dwarfs \citep{Marley96, Burrows97, Marley02}, albeit without incident radiation.  The radiative transfer methods, chemical equilibrium calculations, and molecular and atomic opacities are summarized in the above publications as well as R.~S.~Freedman \& K.~Lodders, in prep.  We assume solar metallicity atmospheres \citep{Lodders03}, and while the effect of condensation on atmospheric composition is included in the chemical equilibrium calculation \citep{Lodders02,Lodders06} we neglect the opacity of clouds.

\begin{figure}
\plotone{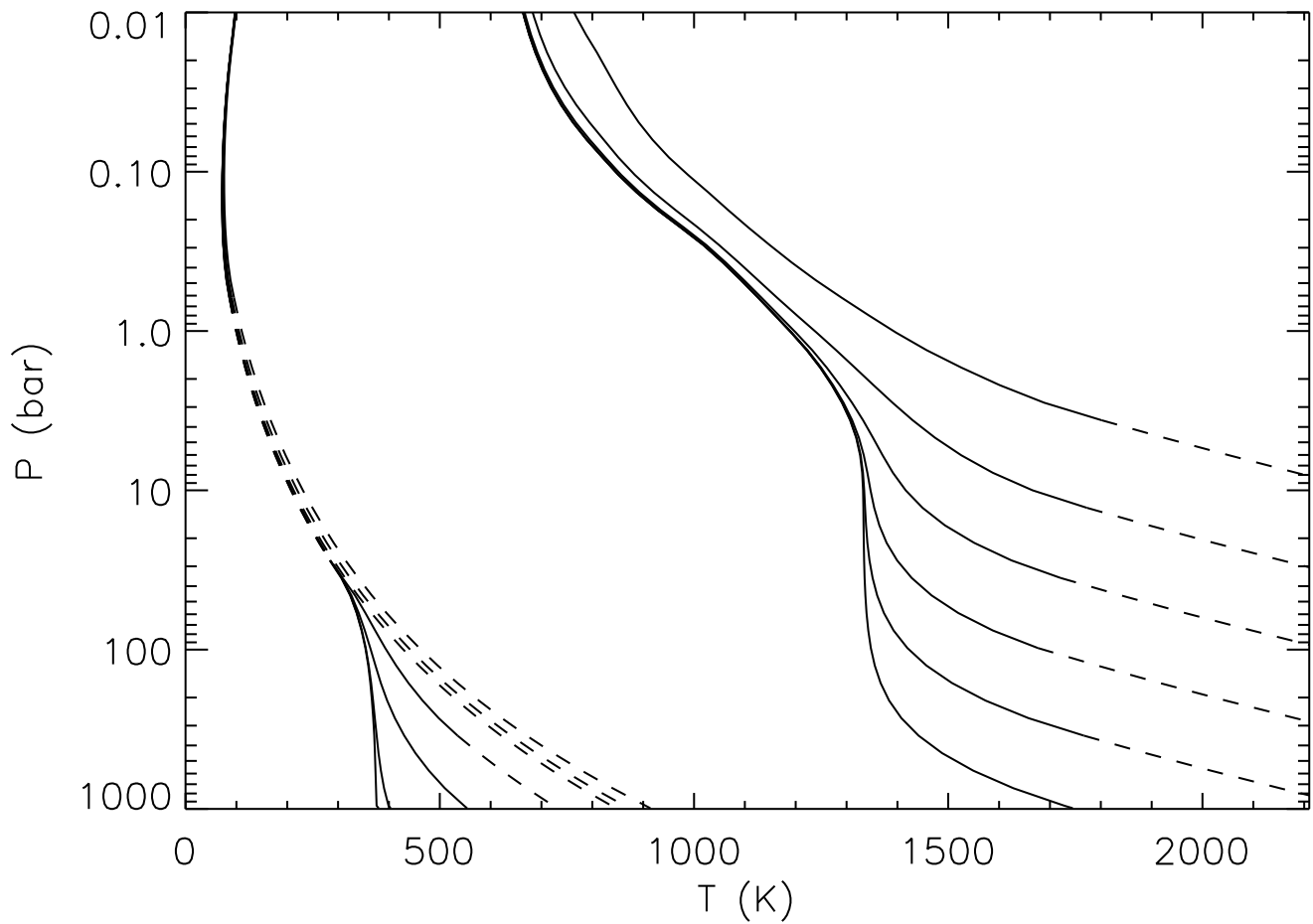}
\caption{A subset of pressure-temperature profiles taken from two of our grids.  On the left are profiles for a planet at 9.5 AU with $g$=13 m s$^{-2}$, decreasing in $T_{\rm{int}}$, with values of 60, 50, 40, 30, 20, 10 and 3 K.  On the right are profiles for a planet at 0.1 AU with $g$=40 m s$^{-2}$, decreasing in $T_{\rm eff}$, with values of 1000, 630, 400, 250, 160, and 100 K.  The solid portions of the profiles are radiative regions and the dashed portions are convective regions.
\label{fig:lowpt}}
\end{figure}

\citet{Guillot02}, \citet{Baraffe03}, \citet{Burrows04}, and \citet{Marley06} have previously discussed the importance of properly incorporating incident stellar flux into the hot Jupiter atmosphere grids that serve as the upper boundary condition for the evolution of these planets, but this is also necessary at greater orbital separations for old and relatively low-mass planets.  These planets have comparatively little internal energy, and their intrinsic effective temperature ($T_{\rm{int}}$) quickly falls below its equilibrium temperature ($T_{\rm {eq}}$), which is set entirely by absorption of stellar flux.  Here, by definition:
\begin{equation}
T_{\rm{eff}}^4 = T_{\rm{int}}^4 + T_{\rm{eq}}^4.
\end{equation}
When $T_{\rm{int}}$ is small, incident stellar flux dominates over intrinsic flux and a deep atmospheric radiative zone grows, similar to a highly irradiated hot Jupiter.  This was previously discussed by J.~B.~Pollack in the early 1990s and mentioned in \citet{Guillot99b}, but we believe this is the first time this effect has been explicitly shown with detailed model atmospheres.  \mbox{Figure~\ref{fig:lowpt}} shows model atmospheres computed at 0.1 AU (highly irradiated) and 9.5 (modestly irradiated).  In both cases once $T_{\rm{int}} \ll T_{\rm {eq}}$, an isothermal region connects the deep interior adiabat to the upper atmosphere, whose structure is governed only by absorption of stellar flux \citep[see][]{Hubeny03}.  Note that the very low $T_{\rm{int}}$ values in \mbox{Figure~\ref{fig:lowpt}} for this Saturn-like planet would only occur after several Hubble times of evolution.  For all models we calculate a planet-wide average \emph{P--T} profile that is representative of the planet as a whole.  In practice this means that the incident stellar flux is diluted by a factor of 1/4 \citep[see][]{Marley06}.

The common approximation described in \citet{Hubbard77} for including stellar flux into an atmosphere grid computed for \emph{isolated} model atmospheres is only valid when $T_{\rm{int}} \gtrsim T_{\rm{eq}}$.  Following the \citet{Hubbard77} prescription to small $T_{\rm{int}}$ leads to planetary radii that reach an asymptotic value governed by their Bond albedo \citep{Chabrier00}, and hence overestimates the radii of old, cold planets.  We have computed model atmosphere grids across a range of surface gravities, for $T_{\rm{int}}$ from 50-2000 K (and as low at 10 K at low gravity), including the proper solar insolation\footnote{The solar spectrum can be dowloaded at http://rredc.nrel.gov/solar/spectra/am0/ASTM2000.html.} for orbital distances of 0.02, 0.045, 0.1, 1.0, and 9.5 AU (which is Saturn's orbital distance).\footnote{Occasionally at higher gravities and low $T_{\rm{int}}$ extrapolation off the grid was performed as well.  This generally only affected highly irradiated core-free low-mass planets.}  At Gyr ages, we find Bond albedos of $\sim$0.05-0.1 at distances less than 0.1 AU, and higher values of $\sim$0.3-0.4 from 1-10 AU, where cooler temperatures prevail and sodium and potassium, which absorb strongly in the optical, have condensed into clouds below the visible atmosphere \citep{Sudar03}.  We have elected to ignore cloud opacity here, since considerable uncertainties remain concerning their effect on the atmospheric structure and albedos of EGPs.  We will pursue this area in more detail in a later paper that focuses on the evolution of Jupiter and Saturn.  A separate important issue is the opacity in the deep atmosphere at pressures near 1 kbar.  While the temperature structure of highly irradiated atmospheres near $P\sim$1 kbar is of great importance for understanding giant planet thermal evolution \citep{Guillot02,Arras06}, the opacities at these pressures remain highly uncertain \citep{depater05}.

\begin{figure}
\plotone{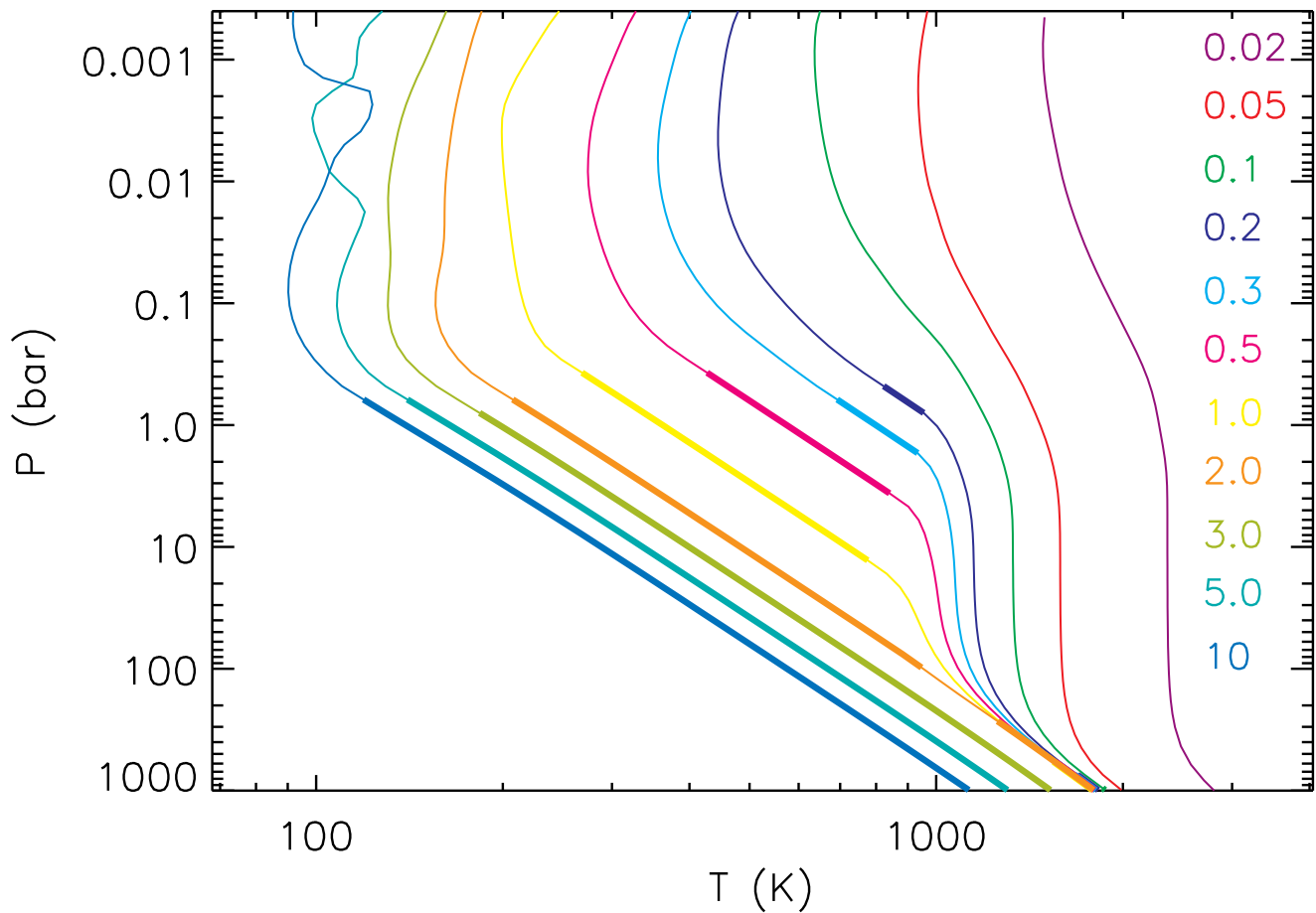}
\caption{Pressure-temperature profiles for $\sim$4.5 Gyr-old Jupiter-like planets ($g$=25 m s$^{-2}$, $T_{\rm{int}}$=100 K) from 0.02 to 10 AU from the Sun.  Distance from the Sun in AU is color coded along the right side of the plot.  Thick lines are convective regions while thin lines are radiative regions.  The profiles at 5 and 10 AU show deviations that arise from numerical noise in the chemical equilibrium table near condensation points, but this has a negligible effect on planetary evolution.
\label{fig:pts}}
\end{figure}

\mbox{Figure~\ref{fig:pts}} shows \emph{P--T} profiles computed for a Jupiter-like planet from 0.02 to 10 AU from the Sun.  The surface gravity $g$ is 25 m s$^{-2}$ and $T_{\rm{int}}=100$ K in all models.  These profiles are meant to roughly illustrate the atmospheres of Jupiter-like planets at 4.5 Gyr.  $T_{\rm{int}}=100$ K is very close to Jupiter's current value \citep{Guillot05}, and we find that in our cooling calculations that the model planets reach a $T_{\rm{int}}$ of $\sim$102-110 K at 4.5 Gyr, which is only weakly dependent on stellar irradiation.

A deep external radiative zone is found in the most highly irradiated models.  For the planets at $\lesssim$0.05 AU convection does not begin until $P > 1$ kbar.  From 0.1 to 2 AU the deep internal adiabat for all models begins at 300+ bar, but there is a 2nd, detached convective zone at pressures close to 1 bar.  This detached convective zone grows at stellar distance increases, and by 3 AU the convective zones have merged.  Only when these convective zones merge is the interior adiabat cooler as a function of orbital distance.  The models from 0.1 to 2 AU have essentially the same internal adiabat, meaning the planets would have the same radius at a given mass.  As we will see, a striking consequence of this effect is that stellar irradiation at 2 AU has approximately the same effect on retarding cooling and contraction as at 0.1 AU, even though the incident fluxes vary by a factor of 400!

\section{Results: Ice--Rock--Iron Planets}
\subsection{Planetary Radii}
It seems likely that planets with masses within an order of magnitude of the Earth's mass will be composed primarily of more refractory species, like the planetary ices, rocks, and iron.  Within our solar system, objects of similar radius can differ by over a factor of 3 in mass, due to compositional differences.  A planet with the radius of Mercury, which is potentially detectable with Kepler, could indicate a mass of 0.055 \me, like Mercury itself, or a mass of 1/3 this value, like Callisto, which has a radius that differs by only 30 km.  With our equations of state, we are able to explore the radii of objects with any possible combination of ice, rock, and iron.  In order to keep this task manageable, we have limited our calculations to several illustrative compositions.  These include pure ice and ice/rock mixtures, which could be described as ``water worlds'' or ``Ocean planets."  Such objects in our solar system, like the icy satellites of the outer planets, generally have small masses.  However, \citet{Kuchner03} and \citet{Leger04} have pointed out water-rich objects could reach many Earth masses (perhaps as failed giant planet cores) and migrate inward to smaller orbital distances.  We also consider planets composed of pure rock, rock and iron mixtures, and pure iron, more similar to our own terrestrial planets.  The ice/rock and rock/iron mixtures are computed for 75/25, 50/50, and 25/75 percentages by mass, with ice always overlaying rock, and rock always overlaying iron.

\begin{figure}
\plotone{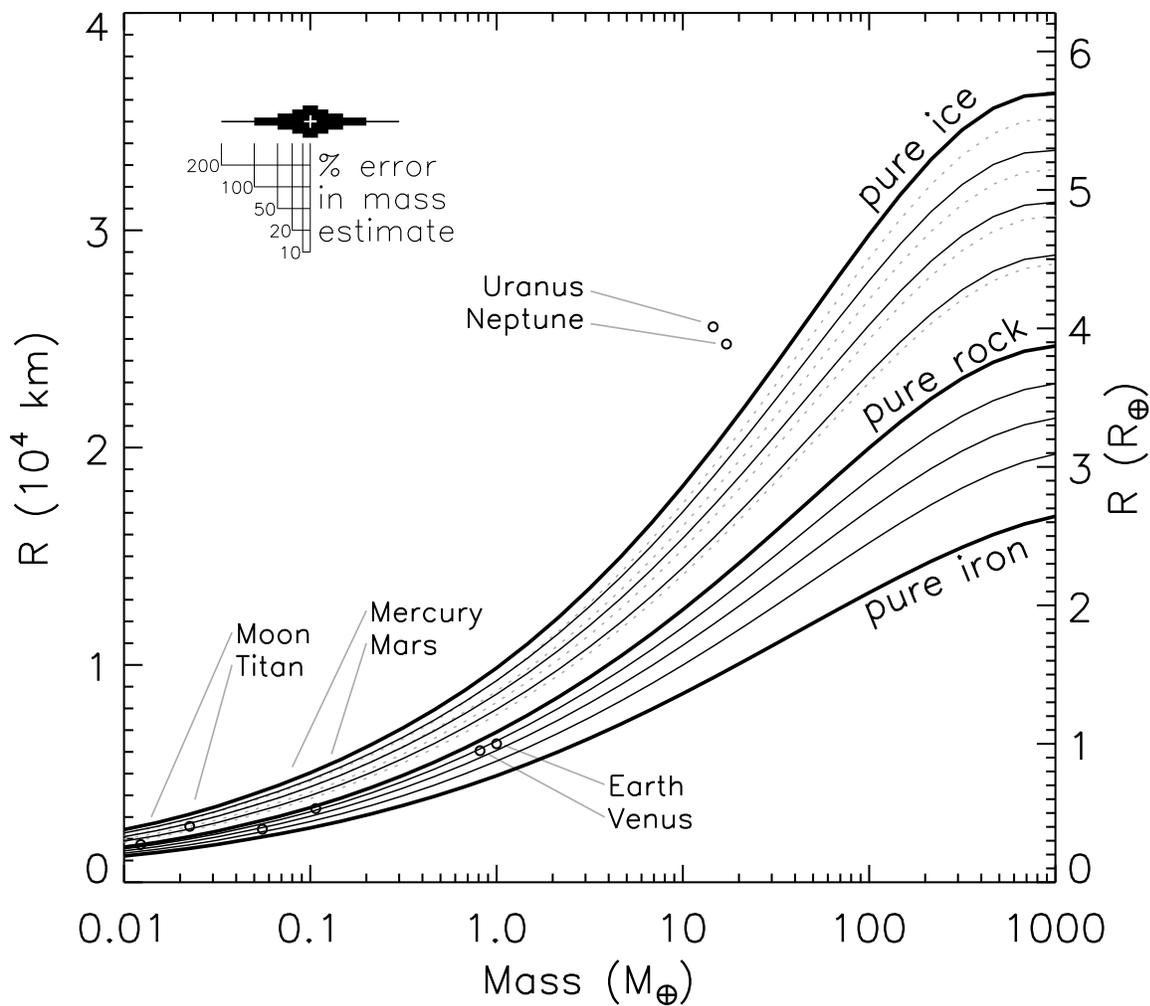}
\caption{Mass (in \me) vs.~radius (in km and \re) for planets composed for ice, rock, and iron.  The topmost thick black curve is for pure ``warm'' water ice.  (See text.)  The middle thick curve is for pure rock (Mg$_2$SiO$_4$).  The bottommost thick curve is for pure iron (Fe).  The three black thin curves between pure ice and pure rock, are from top to bottom, 75\% ice/25\% rock, 50/50, and 25/75.  The inner layer is rock and the outer layer is ice.  The gray dotted lines between rock and pure warm ice are the same pure ice and ice/rock curves, but for zero-temperature ice.  The three black thin curves between pure rock and iron, are from top to bottom, 75\% rock/25\% iron, 50/50, and 25/75.  The inner layer is iron and the outer layer is rock.  Solar system objects are open circles.  At the upper left we show the horizontal extent of mass error bars, for any given mass, from 10 to 200\%.
\label{fig:rmiri}}
\end{figure}

Our results are shown in \mbox{Figure~\ref{fig:rmiri}}.  Since we make few assumptions regarding what is a reasonable planet, we have computed radii from masses of 0.01 to 1000 \me.  For all compositions, the radii initially grow as $M^{1/3}$, but at larger masses, compression effects become important.  As a greater fraction of the electrons become pressure ionized, the materials begin to behave more like a Fermi gas, and there is a flattening of the mass-radius curves near 1000 \me.  Eventually the radii shrink as mass increases, with radii falling with $M^{-1/3}$ \citep[see][]{Zapolsky69}.
 
At the top left of \mbox{Figure~\ref{fig:rmiri}} we also show the size of various levels of uncertainty in planetary mass, as a percentages of a given mass, from 10 to 200\%.  For instance, if one could determine the mass of a 1 \me~planet to within 50\%,  even a radius determination accurate to within 0.25 \re~would lead to considerable ambiguity concerning composition, ranging from 50/50 ice/rock to pure iron.  The shallow slope of the mass-radius curves below a few \me~makes accurate mass determinations especially important for understanding composition.  In \mbox{Table~\ref{tiri}} we give the mass and radius for a subset of these planets.  We note that from 1-10 \me~we find excellent agreement between our models and the more detailed ``Super-Earth'' models of \citet{Valencia06}.

\subsection{Validation in the Solar System}
On \mbox{Figure~\ref{fig:rmiri}} we have also plotted, in open circles, the masses and radii of solar system planets and moons.  These planets can be used to validate our methods.  For instance, detailed models of the Earth's interior indicate that the Earth is approximately 33\% iron by mass with a core-mantle boundary at 3480 km \citep{prem}.  This composition is readily recovered from \mbox{Figure~\ref{fig:rmiri}}, where Earth plots between the 25\% and 50\% iron curves, but closer to 25\%.  Our simple Earth model, with a iron/rock boundary at 3480 km yields a planetary radius within 100 km (1.5\% smaller) of the actual Earth.  Given that our model lacks thermal corrections to EOSs that are found in detailed Earth models and that we ignore lower density species such as sulfur that are likely mixed with iron into the Earth's core, we regard this agreement as excellent, and entirely sufficient with regard to the expected radii uncertainties as measured by transit surveys.

Elsewhere in the solar system, one can see that we recover ice/rock or rock/iron ratios of other bodies, which are derived by more complex models.  A brief overview of the structure of the terrestrial planets and icy moons is given in \citet{dePater01}.  Earth's moon is composed almost entirely of rock, with a very small iron core of radius $\lesssim 400$ km.  Here, the moon (the leftmost circle) plots on top of the line for pure rock.  Mercury is calculated to be $\sim$60\% iron by mass, and with our models Mercury falls between the 50/50 (rock/iron) and 25/75 curves, but again, closer to 50/50, which shows excellent agreement.  Titan is calculated to be composed of $\sim$35\% ices, and again we find excellent agreement, as Titan falls between the 50/50 (ice/rock) and 25/75 curves, slightly closer to 25/75.  Uranus and Neptune cannot be composed purely of ice, although ice likely makes up the bulk of their masses.

\subsection{Applications}
In addition to \mbox{Table~\ref{tiri}} we have also fit the radii of these planets to analytic functions that are quadratic in $\log{M}$ and linear in composition.  These are Equation (\ref{fit1}), for ice/rock planets, and (\ref{fit2}), for rock/iron, shown below:

\begin{eqnarray}
\label{fit1}
R=(0.0912 \quad{imf} + 0.1603)(\log{M})^2 + (0.3330 \quad{imf} + 0.7387)\log{M} \nonumber\\ + (0.4639 \quad{imf} + 1.1193).\quad \quad \quad \quad
\end{eqnarray}
\begin{eqnarray}
\label{fit2}
R=(0.0592 \quad{rmf} + 0.0975)(\log{M})^2 + (0.2337 \quad{rmf} + 0.4938)\log{M} \nonumber\\ + (0.3102 \quad{rmf} + 0.7932) \quad \quad \quad \quad
\end{eqnarray}

Here $R$ is in \re~and $M$ is in \me, while ${imf}$ is the ice mass fraction (1.0 for pure ice and 0.0 for pure rock) and ${rmf}$ is the rock mass fraction (1.0 for pure rock and 0.0 for pure iron).  The fits were performed for planetary masses from 0.01 to 100 \me.  For ice/rock, Equation (\ref{fit1}) is on average accurate to within 2.5\%.  For rock/iron, Equation (\ref{fit2}) is accurate to 1.5\%.  Accuracy near 1 \me~for both equations is better than 0.5\%, although deviations can reach 10\% for ice/rock at $\sim$0.01 \me.  If a mass and radius can be determined for a given planet these equations will allow for a quick and reliable composition estimate.  However, since radial velocity or astrometric followup for these small planets will be extremely difficult and time consuming, radii may have to suffice as a proxy for mass for some time.  Given an assumed composition, such as ``Earth-like," one could assign masses to terrestrial-sized transiting planets.  The distribution of planetary masses vs.~orbital distance and stellar type could then be compared with planet formation models.

\section{Results: Hydrogen-Helium Dominated Planets}
\subsection{Radii of Gas Giants}
\begin{figure}
\plotone{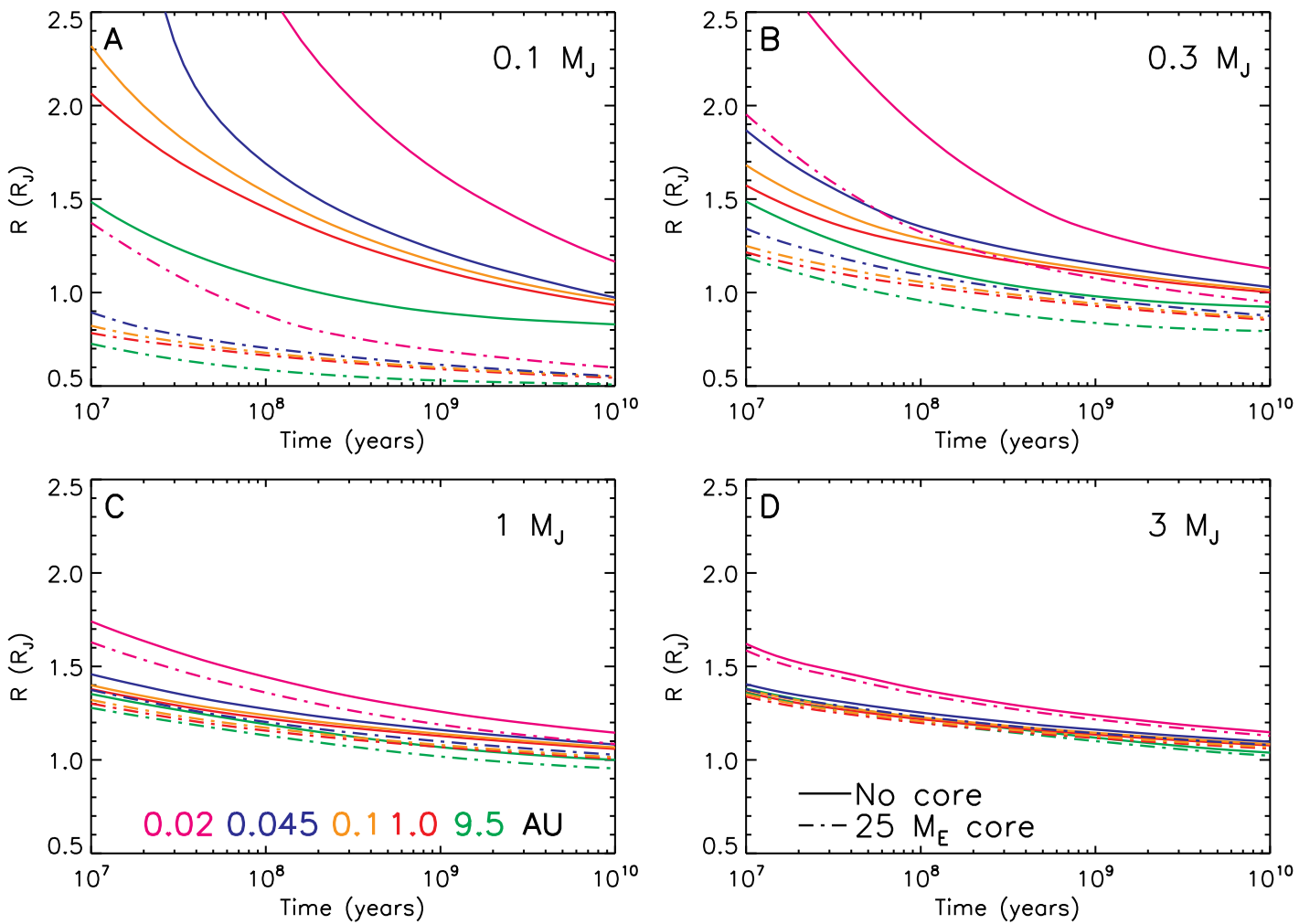}
\caption{Planetary radii as a function of time for masses of 0.1 \mj~(32 \me, $A$), 0.3 \mj~($B$), 1.0 \mj~($C$), and 3.0 \mj~($D$).  The five curve colors code for the five different orbital separations from the Sun, shown in ($C$).  Solid lines indicate models without cores and dash-dot lines indicate models with a core of 25 \me.
\label{fig:quad}}
\end{figure}

Planets around the mass of Uranus and Neptune ($\sim15$ \me) to objects as large at 75 \mj~can be described by the same cooling theory.  In general, planets with larger cores will have smaller radii, and planets closer to their parent stars will have larger radii at a given age then planets at larger orbital distances.  In \mbox{Figure~\ref{fig:quad}} we plot the contraction of planets from 0.1 to 3 \mj, as a function of age at various orbital distances.  We also show the effect of a core of 25 \me, the approximate mass of heavy elements within Jupiter and Saturn \citep{Saumon04}.  The first feature to notice is that, independent of mass, the spread in radii between 0.045 and 1 AU is quite small compared to the large radii at 0.02 AU and small radii at 9.5 AU.  As expected, the effect of the 25 \me~core diminishes with increased planet mass, as the core becomes a relatively smaller fraction of the planet's mass.  The radii at early ages it quite large, especially for the low-mass planets under intense stellar irradiation.  These hot low-gravity planets are potentially susceptible to evaporation \citep{Baraffe05, Hubbard07}.

As discussed in detail by \citet{Marley07}, the physical properties of giant planets at young ages are quite uncertain.  The models presented here do not include a formation mechanism and are arbitrarily large and hot at very young ages.  \citet{Marley07} have found that in their implementation of the core-accretion mechanism of giant planet formation \citep{Pollack96, Hubickyj05} giant planets form with radii than can be several tenths of a Jupiter radius smaller than one computes with an arbitrarily large and hot start.  These differences may last $\sim$10 Myr for a 1 \mj~object to hundreds of millions of years for planets of several Jupiter masses.  Therefore, determining the radius of young giant planets, has the potential to elucidate their post-formation structure, and give us clues to their formation mechanism.  In addition, a young hot Jupiter would place constraints on planetary migration times \citep{Burrows00}.

\begin{figure}
\plotone{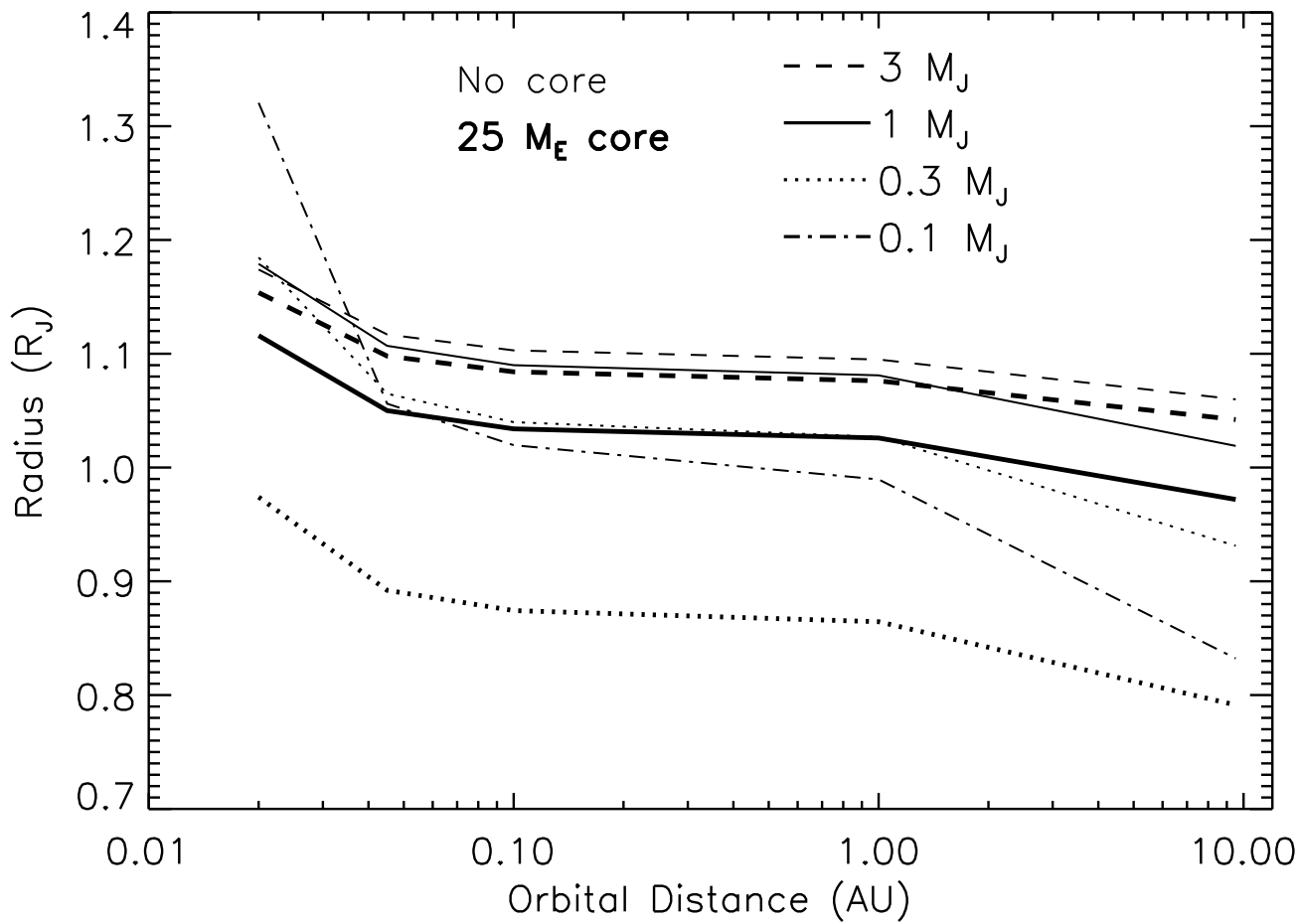}
\caption{Planetary radii at 4.5 Gyr as a function of orbital distance from the Sun.  Models are calculated at 0.02, 0.045, 0.1, 1.0, and 9.5 AU.  Masses are 0.1, 0.3, 1.0, and 3.0 \mj.  Coreless planets (thin lines) and planets with a core of 25 \me~of heavy elements (thick lines) are shown.  Note the shape of these radius curves and the flattening between 0.1 and 1.0 AU.  The 0.1 \mj~planet with a 25 \me~core is off the plot at $\sim$0.5 \rj.
\label{fig:rd}}
\end{figure}

Radii as a function of orbital distance, at 4.5 Gyr, are shown in \mbox{Figure~\ref{fig:rd}}.  For the planets from 0.3 (Saturn's mass) to 3 \mj, the radius curves are nearly flat between 0.1 and 1 AU.  These models predict that irradiation effects will remain important for planets $>0.1$ AU from their parent stars.  \emph{Planets with the same composition, from 0.1 to $\sim$2 AU, should have nearly the same radii.}  This is a consequence of the atmospheric temperature structures shown in \mbox{Figure~\ref{fig:pts}}.  Planets with cores are smaller, but there radii curves also flatten from 0.1 to $\sim$1 AU.

\begin{figure}
\plotone{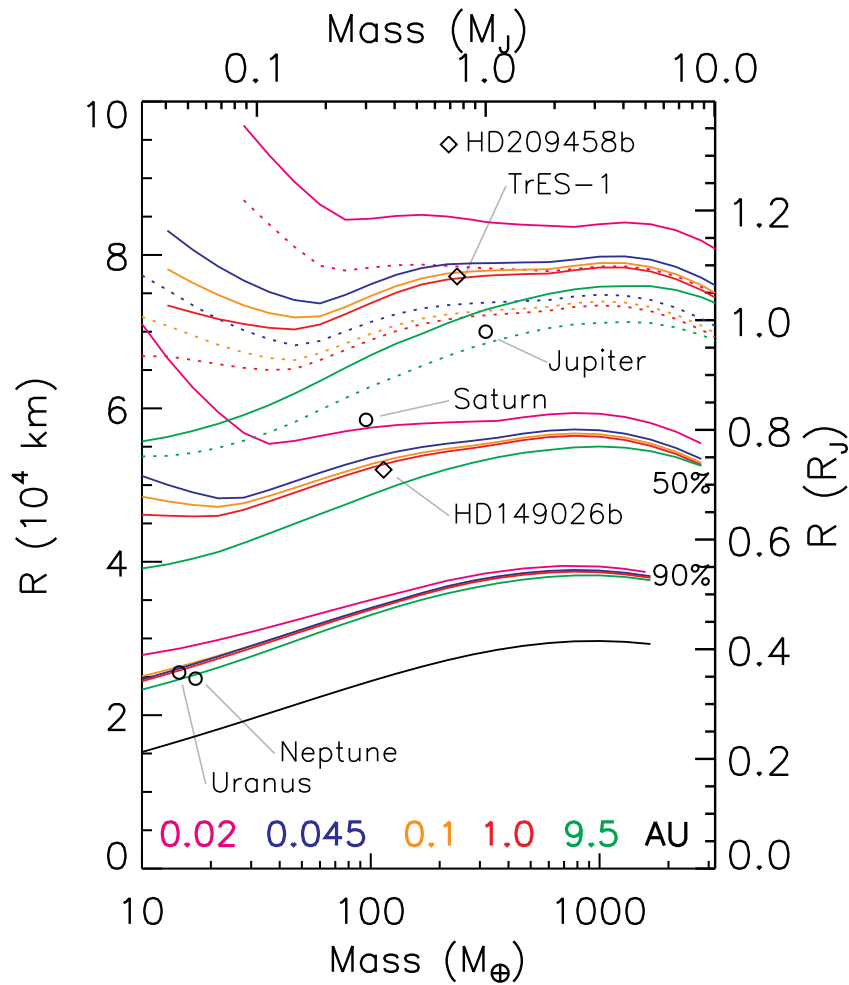}
\caption{Planetary radii at 4.5 Gyr as a function of mass.  Models are calculated at 0.02, 0.045, 0.1, 1.0, and 9.5 AU and are color coded at the bottom of the plot.  The black curve is for a heavy element planet of half ice and half rock.  The group of 5 colored curves above the black curve is for planets that are 90\% heavy elements.  The next higher set of 5 colored curves are for planets that are 50\% heavy elements.  The next higher set, shown in dotted lines, are 10\% heavy elements.  The highest set are for core-free planets of pure H/He.  The open circles are solar system planets and the diamonds are extrasolar planets. 
\label{fig:rmhhe}}
\end{figure}

We can also consider the effects of extremely large and small core masses.  One view of our full range of 4.5 Gyr-old planets is shown in \mbox{Figure~\ref{fig:rmhhe}} where radii vs.~mass is plotted at 5 orbital separations from 0.02 AU to 9.5 AU.  Coreless planets and planets with cores that are 10\%, 50\%, and 90\% of their \emph{total planet mass} are shown.  Note that these models have a core mass mass that is a constant \emph{percentage} of planetary mass, meaning that a 50\% core mass for a 10 \me~planet is 5 \me, and for a 5 \mj~planet it is 800 \me.  This plot is meant to show extremes, but is still quite illustrative.  Several real planets are shown as well.

In \mbox{Figure~\ref{fig:rmhhe}} Uranus and Neptune lie very close to the 90\% heavy elements curves, as expected from more detailed models \citep{Hubbard91,Podolak95}, although these planets likely contain more ice than rock.  The radii of ``hot Neptunes,'' should be similar to that of Uranus and Neptune, \emph{if} hot Neptunes do not suffer the effects of significant mass loss and are $\sim$10\% H/He, like Uranus and Neptune.  \citet{Baraffe05} and \citet{Baraffe06} have shown that if these planets are remnants of much larger original planets, their radii may well exceed 1 \mj.

Looking at the solar system's gas giants, it is clear that for Jupiter\footnote{Jupiter itself plots below 1 \rj~because its mean radius is 2.2\% smaller than its equatorial radius at 1 bar, 71492 km, which has become the standard ``Jupiter radius.''  We note that planetary oblateness will be extremely difficult to determine from transit light curves \citep{Barnes03,Seager02}.} at 5.2 AU, 10\% heavy elements by mass is a reasonable estimate as it falls between the 1 AU and 9.5 AU 10\% heavy element curves.  Saturn, at 9.5 AU, is clearly more enhanced in heavy elements that Jupiter, and from this plot perhaps 20\% heavy elements would be estimated.  Given that one of the first ten transiting planets found has a core mass of $\sim$60-80 \me, and it orbits a star with a metallicity 2.3$\times$ that of the Sun, it is probably realistic to expect the occasional planet around metal rich stars with core masses of $\sim$100 \me~or more. 

\begin{figure}
\plotone{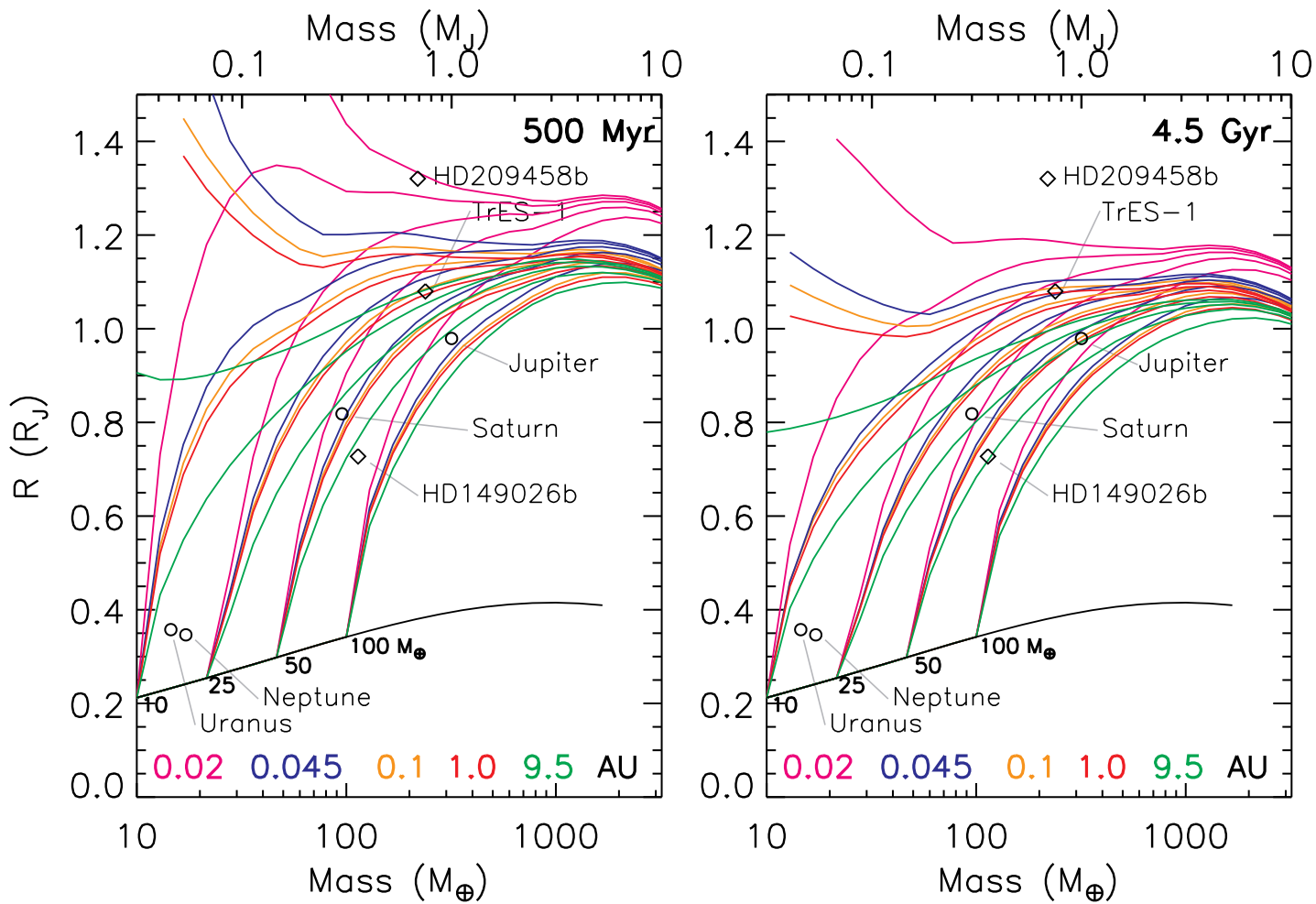}
\caption{Planetary radii with various core masses at 500 Myr ($A$) and 4.5 Gyr ($B$).  Models are calculated at 0.02, 0.045, 0.1, 1.0, and 9.5 AU and are color coded at the bottom of the plot.  The black curve is for a heavy element planet of 50\% ice and 50\% rock.  Models with no core, and core masses of 10, 25, 50, and 100 \me~are computed.  Labels in thick black text show where curves with a given constant core mass fall upon the mass/radius curve for heavy elements.  Planets at 0.045, 0.1, and 1.0 AU are similar in radius at every mass, but planets at 0.02 AU (magenta) are significantly larger while planets at 9.5 AU (green) are significantly smaller.  Core-free models are the curves that terminate at low mass at the upper left.  The open circles are solar system planets and the diamonds are extrasolar planets.
\label{fig:rcores}}
\end{figure}

We now turn to models computed with a constant core mass (10, 25, 50, and 100 \me) as function of age, mass, and stellar irradiation.  \mbox{Figures~\ref{fig:rcores}\emph{a}} and \ref{fig:rcores}\emph{b} are dense plots that show the radii of planets at two ages, 500 Myr and 4.5 Gyr.  The contraction of planets by $\sim$0.1-0.2 \rj~across this factor of $\sim$10 in age is clear.  Planets at 0.045 AU (blue), 0.1 AU (orange), and 1.0 AU (red) are very similar in radius at every mass, so their mass-radius curves trace similar path, while planets at 0.02 AU (magenta) are always substantially larger, and planets at 9.5 AU (green) are always substantially smaller.  In Tables~\ref{t300}, \ref{t1000}, and \ref{t4500} contain our calculations of the radii of these planets at ages of 300 Myr, 1 Gyr, and 4.5 Gyr.  Looking at radii at 4.5 Gyr, we find generally good agreement with the models of \citet{Bodenheimer03} for planets $\gtrsim$0.69 \mj~(within 0.03 \rj), but at lower masses we predict significantly smaller planets with differences than can approach 0.4 \rj~for coreless models at 0.02 AU.  It is likely that in the very low-mass core-free, highly irradiated corner of phase space that evolution models are perhaps most uncertain.  Evaporation effects may be important here as well.

\subsection{Applications}
Extending our calculations to parent stars other than the Sun should be done with some care.  \citet{Marley99} have shown that planetary Bond albedos are a function of the spectral type of the primary star.  A planet around a later-type star, with a spectrum peaking closer to the infrared, will have a greater fraction of its flux absorbed by a planetary atmosphere, meaning a lower planetary Bond albedo.  Although this must surely be accounted for eventually, given current uncertainties in chemistry and cloud formation in these planets, which leads directly to uncertainties in their atmospheric absorption and scattering properties, for now we will ignore this effect.  However, the incident stellar fluxes of the currently known transiting planets already differ by a factor of \emph{18}\footnote{See Frederic Pont's website  http://obswww.unige.ch/~pont/TRANSITS.htm for an updated tabulation of transiting planet system parameters.}, so the differing magnitudes of incident radiation cannot be ignored.  If a planet's orbital distance from its parent star is $d$, then the distance from the Sun that the planet would have to be to receive this same flux, $d_{\odot}$, is given by:
\begin{equation}
d_{\odot}=d \left( \frac{L_{\odot}}{L} \right) ^{1/2}
\end{equation}
or, if one uses a mass-radius relation such as,
\begin{equation}
\frac{L}{L_{\odot}}=\left( \frac{M}{M_{\odot}} \right) ^{\eta},
\end{equation}
where $\eta$ may typically be $\sim$3.5 across the H-R diagram, then
\begin{equation}
d_{\odot}=d \left( \frac{M_{\odot}}{M} \right) ^{\eta/2}.
\end{equation}
For instance, \hd, at 0.046 AU from its G0V parent star ($L\sim1.6$ $L_{\odot}$), would receive that same level of irradiation at 0.036 AU from the Sun.

\mbox{Figure~\ref{fig:rcores}} shows the mass and radius of three transiting planets.  The planet with the lowest mass, \hh, receives an incident flux equivalent to that of a planet at 0.026 AU from the Sun.  The figure clearly suggests that the planet has a core of perhaps 70 \me, which is confirmed by more detailed models \citep{Fortney06}.  The middle diamond is \T, which receives insolation equal to what is received from the Sun at 0.059 AU.  Using these models, one would estimate that the planet has a core of perhaps 25~\me, very similar to the bulk abundance of heavy elements in Jupiter and Saturn \citep{Saumon04}.  \hd~is anomalous on this plot and must have an additional interior energy source.  However, a 0.1 \mj~planet of pure H/He at 0.02 AU (if such a planet exists) with a similar radius would have the correct radius.

It should also be pointed out that all the hot Jupiters could have additional energy sources of similar magnitude.  As discussed in \citet{Fortney06} and \citet{Guillot06}, this would require planets such as \T~and \hh~to have even larger cores.  With a common energy source, differences in the radii of giant planets would then be attributable to differing masses of heavy elements within their interiors.  This idea is strengthened by the recent discovery of hot Jupiters with large radii, such as HAT-P-1b \citep{Bakos06} and WASP-1b \citep{Collier07, Charb07}, which shows that inflated radii are common features of these irradiated planets.

\section{Discussion \& Conclusions}
In this paper we have taken a broad look at the radii of planets at nearly all possible masses and orbital separations.  These calculations serve as a baseline for comparisons with transiting planets currently being discovered from the ground and those that will soon be discovered from space.  Since our main effort here is to calculate radii over a very large phase space, there are surely particular corners of space where these calculations could be improved, as we outline below.  If in the coming years it becomes clear under what conditions planets may have an additional interior energy source, and how this changes as a function of orbital separation, incident stellar flux, and stellar type, this could be included in a new grid of calculations.  If evaporation effects are observed for numerous transiting planets, across a wide range of planetary surface gravities and orbital separations, then the effects of appreciable mass loss could be added to these evolution calculations.  The work of \citet{Baraffe06} and \citet{Hubbard07} is moving in this direction.  

Ice-rich terrestrial planets close to their parent stars could have steam atmospheres \citep{Kuchner03} which, depending on the atmospheric scale height, could lead to larger radii at a given mass for these objects.  If the cooling of Uranus and Neptune are eventually better understood, then the additional energy due to the cooling of their heavy element interiors could be added to our evolution models of these lower-mass EGPs.  Including the EOS of other forms of rock, or other astrophysical ices, such at CH$_4$ and NH$_3$, would be useful when more detailed models are eventually needed.  One could also envision more exotic planets, such as those that formed in solar systems where C/O$>$1, which could lead to carbon dominated ``terrestrial planets" \citep{Kuchner05}, which would likely have radii intermediate between pure rock and ice (Zapolsky \& Salpeter, 1969; M.~Kuchner, personal communication).

As there is in the solar system, there will always be ambiguity in the bulk composition of exoplanets, especially those that do not posses substantial low density gaseous envelopes.  For instance, since a mixture of H/He gas and rocks has an EOS that is similar to that of ice, this leads to considerable uncertainty in the composition of Uranus and Neptune.  For a terrestrial-mass ``ocean planet,'' perhaps the best that we can expect will be be an understanding of a planet's ice/rock ratio, with a given uncertainty based on the mass and radius measurements, along with the knowledge that a the uncertainty could be even larger if the planet has a substantial iron core.  Since iron and silicates condense at similar temperatures \citep{Lodders03}, it is possible that these species will condense out in ratios similar to that found in the Earth, although subsequent collisions (e.g., as experienced by the Moon and probably Mercury) could alter this ratio, especially for smaller bodies.  Experience will tell us how to best classify newly discovered terrestrial-type planets, but at this point we would advocate composition classes based on simple ratios of ice/rock and rock/iron.  Our analytic fits reproduce the behavior of our models and will allow for mass estimates prior to radial velocity or astrometric follow up work.

A clear prediction from our models for giant planets is that irradiation effects on planetary radii are not a simple function of stellar insolation.  At a given planetary mass and composition, we predict a flattening in planetary radius as a function of orbital distance from $\sim$0.1 to $\sim$1-2 AU.  However, if unknown additional energy sources are still in play at these orbital distances, then this effect may be more difficult to see.

Given the recent flurry of transit detections it is likely that we are now just seeing the tip of the iceberg of transiting planets.  Even before detections from COROT and Kepler are announced it seems likely that we will have a steady stream of new planets to challenge our understanding.

\acknowledgements
We thank Kevin Zahnle for a helpful discussion at the start of this project.  JJF acknowledges the support of a Spitzer Fellowship from NASA and NSF grant AST-0607489 and MSM from the NASA Origins and Planetary Atmospheres Programs.

%\bibliographystyle{apj}
%\bibliography{references}

\clearpage

\begin{deluxetable}{ccccccccccc}
\center
\tablecolumns{11}
\tablewidth{0pc}
\tablecaption{Ice-Rock-Iron Planetary Radii}
\tablehead{
\colhead{Composition} & \colhead{0.010} & \colhead{0.032} & \colhead{0.1} & \colhead{0.32} & \colhead{1.0} & \colhead{3.16} & \colhead{10.0} & \colhead{31.6} & \colhead{100} & \colhead{316}}
\startdata 
ice & 0.38 & 0.55 & 0.79 & 1.12 & 1.55 & 2.12 & 2.87 & 3.74 & 4.68 & 5.43\\ 
50/50& 0.33 & 0.48 & 0.69 & 0.97 & 1.36 & 1.85 & 2.48 & 3.23 & 4.03 & 4.67\\ 
rock & 0.25 & 0.37 & 0.54 & 0.77 & 1.08 & 1.48 & 1.97 & 2.54& 3.14 & 3.64\\ 
67/33 (Earth-like)& 0.24 & 0.34 & 0.50 & 0.71 & 1.00 & 1.36 & 1.80 & 2.31& 2.84 & 3.29\\
50/50& 0.23 & 0.33 & 0.48 & 0.68 & 0.95 & 1.30 & 1.71 & 2.19& 2.69 & 3.12\\
iron & 0.19 & 0.27 & 0.39 & 0.55 & 0.77 & 1.04 & 1.36 & 1.72& 2.09 & 2.42\\
\enddata
\tablecomments{Radii of planets, in \re.  Column headers are planet masses, in \me.}
\label{tiri}
\end{deluxetable}

\begin{deluxetable}{cccccccccccccc}
\tabletypesize{\footnotesize}
%\rotate
\center
\tablecolumns{14}
\tablewidth{0pc}
\tablecaption{Giant Planet Radii at 300 Myr}
\tablehead{
\colhead{Distance} & \colhead{Core} & \colhead{0.052} & \colhead{0.087} & \colhead{0.15} & \colhead{0.24} & \colhead{0.41} & \colhead{0.68} & \colhead{1.0} & \colhead{1.46} & \colhead{2.44} & \colhead{4.1} & \colhead{6.8}  & \colhead{11.3} \\ \colhead{(AU)} & \colhead{Mass} & \colhead{17} & \colhead{28} & \colhead{46} & \colhead{77} & \colhead{129} & \colhead{215} & \colhead{318} & \colhead{464} & \colhead{774} & \colhead{1292} & \colhead{2154}  & \colhead{3594} }
\startdata

  0.02 &   0 & * & 2.326 & 1.883 & 1.656 & 1.455 & 1.378 & 1.342 & 1.327 & 1.308 & 1.311 & 1.315 & 1.284\\
  0.02 &  10 & 1.102 & 1.388 & 1.465 & 1.422 & 1.349 & 1.325 & 1.311 & 1.306 & 1.295 & 1.304 & 1.310 & 1.281\\
  0.02 &  25 & - & 0.493 & 0.945 & 1.133 & 1.220 & 1.253 & 1.267 & 1.275 & 1.276 & 1.294 & 1.304 & 1.277\\
  0.02 &  50 & - & - & - & 0.801 & 1.030 & 1.144 & 1.193 & 1.226 & 1.245 & 1.276 & 1.292 & 1.270\\
  0.02 & 100 & - & - & - & - & 0.669 & 0.939 & 1.055 & 1.128 & 1.187 & 1.242 & 1.270 & 1.256\\
 0.045 &   0 & 2.795 & 1.522 & 1.345 & 1.255 & 1.240 & 1.228 & 1.212 & 1.206 & 1.199 & 1.210 & 1.203 & 1.170\\
 0.045 &  10 & 0.801 & 1.012 & 1.091 & 1.124 & 1.168 & 1.185 & 1.185 & 1.188 & 1.188 & 1.204 & 1.199 & 1.168\\
 0.045 &  25 & - & 0.447 & 0.793 & 0.968 & 1.071 & 1.124 & 1.147 & 1.161 & 1.173 & 1.195 & 1.193 & 1.164\\
 0.045 &  50 & - & - & - & 0.719 & 0.921 & 1.033 & 1.084 & 1.119 & 1.148 & 1.179 & 1.183 & 1.157\\
 0.045 & 100 & - & - & - & - & 0.627 & 0.863 & 0.968 & 1.036 & 1.101 & 1.148 & 1.163 & 1.146\\
   0.1 &   0 & 1.595 & 1.395 & 1.270 & 1.197 & 1.202 & 1.198 & 1.187 & 1.182 & 1.178 & 1.189 & 1.178 & 1.144\\
   0.1 &  10 & 0.755 & 0.956 & 1.035 & 1.084 & 1.134 & 1.157 & 1.160 & 1.164 & 1.168 & 1.183 & 1.174 & 1.142\\
   0.1 &  25 & - & 0.438 & 0.767 & 0.938 & 1.042 & 1.099 & 1.123 & 1.138 & 1.153 & 1.174 & 1.169 & 1.138\\
   0.1 &  50 & - & - & - & 0.702 & 0.899 & 1.011 & 1.063 & 1.098 & 1.129 & 1.158 & 1.159 & 1.132\\
   0.1 & 100 & - & - & - & - & 0.618 & 0.847 & 0.950 & 1.018 & 1.084 & 1.128 & 1.140 & 1.121\\
   1.0 &   0 & 1.504 & 1.325 & 1.222 & 1.169 & 1.182 & 1.182 & 1.173 & 1.169 & 1.168 & 1.179 & 1.169 & 1.136\\
   1.0 &  10 & 0.727 & 0.921 & 1.004 & 1.063 & 1.116 & 1.141 & 1.146 & 1.152 & 1.158 & 1.173 & 1.165 & 1.134\\
   1.0 &  25 & - & 0.433 & 0.754 & 0.923 & 1.027 & 1.085 & 1.110 & 1.127 & 1.143 & 1.164 & 1.159 & 1.130\\
   1.0 &  50 & - & - & - & 0.693 & 0.888 & 0.999 & 1.051 & 1.087 & 1.120 & 1.149 & 1.149 & 1.124\\
   1.0 & 100 & - & - & - & - & 0.613 & 0.839 & 0.941 & 1.009 & 1.075 & 1.119 & 1.131 & 1.113\\
   9.5 &   0 & 0.929 & 0.951 & 0.983 & 1.020 & 1.070 & 1.106 & 1.127 & 1.146 & 1.167 & 1.169 & 1.156 & 1.130\\
   9.5 &  10 & 0.565 & 0.733 & 0.847 & 0.939 & 1.016 & 1.072 & 1.104 & 1.131 & 1.157 & 1.163 & 1.152 & 1.127\\
   9.5 &  25 & - & 0.394 & 0.664 & 0.826 & 0.942 & 1.024 & 1.073 & 1.146 & 1.142 & 1.153 & 1.146 & 1.124\\
   9.5 &  50 & - & - & - & 0.635 & 0.823 & 0.951 & 1.020 & 1.072 & 1.119 & 1.137 & 1.137 & 1.118\\
   9.5 & 100 & - & - & - & - & 0.587 & 0.810 & 0.920 & 0.999 & 1.072 & 1.107 & 1.119 & 1.107\\

\enddata
\tablecomments{Radii of planets, in \rj.  Row 1 column headers are planet masses, in \mj, while row 2 is in \me.  The symbol ``-'' indicates that the planet mass is smaller than the given core mass.}
\label{t300}
\end{deluxetable}

\begin{deluxetable}{cccccccccccccc}
\tabletypesize{\footnotesize}
%\rotate
\center
\tablecolumns{14}
\tablewidth{0pc}
\tablecaption{Giant Planet Radii at 1 Gyr}
\tablehead{
\colhead{Distance} & \colhead{Core} & \colhead{0.052} & \colhead{0.087} & \colhead{0.15} & \colhead{0.24} & \colhead{0.41} & \colhead{0.68} & \colhead{1.0} & \colhead{1.46} & \colhead{2.44} & \colhead{4.1} & \colhead{6.8}  & \colhead{11.3} \\ \colhead{(AU)} & \colhead{Mass} & \colhead{17} & \colhead{28} & \colhead{46} & \colhead{77} & \colhead{129} & \colhead{215} & \colhead{318} & \colhead{464} & \colhead{774} & \colhead{1292} & \colhead{2154}  & \colhead{3594} }
\startdata

  0.02 &   0 & * & 1.770 & 1.539 & 1.387 & 1.309 & 1.281 & 1.258 & 1.248 & 1.235 & 1.244 & 1.240 & 1.199\\
  0.02 &  10 & 0.909 & 1.150 & 1.221 & 1.211 & 1.228 & 1.234 & 1.229 & 1.229 & 1.224 & 1.237 & 1.235 & 1.197\\
  0.02 &  25 & - & 0.461 & 0.838 & 1.022 & 1.121 & 1.169 & 1.189 & 1.200 & 1.206 & 1.228 & 1.229 & 1.192\\
  0.02 &  50 & - & - & - & 0.746 & 0.958 & 1.072 & 1.122 & 1.156 & 1.180 & 1.211 & 1.218 & 1.186\\
  0.02 & 100 & - & - & - & - & 0.640 & 0.888 & 0.997 & 1.068 & 1.130 & 1.179 & 1.198 & 1.173\\
 0.045 &   0 & 1.490 & 1.271 & 1.183 & 1.144 & 1.163 & 1.167 & 1.160 & 1.157 & 1.156 & 1.164 & 1.149 & 1.107\\
 0.045 &  10 & 0.698 & 0.888 & 0.975 & 1.043 & 1.099 & 1.127 & 1.134 & 1.140 & 1.147 & 1.158 & 1.145 & 1.105\\
 0.045 &  25 & - & 0.426 & 0.739 & 0.908 & 1.012 & 1.072 & 1.099 & 1.115 & 1.132 & 1.149 & 1.140 & 1.101\\
 0.045 &  50 & - & - & - & 0.684 & 0.877 & 0.988 & 1.041 & 1.077 & 1.109 & 1.134 & 1.130 & 1.095\\
 0.045 & 100 & - & - & - & - & 0.607 & 0.831 & 0.932 & 0.999 & 1.065 & 1.105 & 1.111 & 1.084\\
   0.1 &   0 & 1.298 & 1.197 & 1.127 & 1.105 & 1.133 & 1.143 & 1.139 & 1.138 & 1.139 & 1.147 & 1.130 & 1.087\\
   0.1 &  10 & 0.665 & 0.847 & 0.934 & 1.012 & 1.072 & 1.105 & 1.114 & 1.122 & 1.130 & 1.141 & 1.126 & 1.085\\
   0.1 &  25 & - & 0.420 & 0.719 & 0.883 & 0.989 & 1.051 & 1.080 & 1.097 & 1.116 & 1.132 & 1.121 & 1.081\\
   0.1 &  50 & - & - & - & 0.670 & 0.859 & 0.970 & 1.023 & 1.059 & 1.094 & 1.117 & 1.111 & 1.076\\
   0.1 & 100 & - & - & - & - & 0.600 & 0.818 & 0.918 & 0.984 & 1.050 & 1.088 & 1.093 & 1.065\\
   1.0 &   0 & 1.229 & 1.148 & 1.095 & 1.086 & 1.118 & 1.130 & 1.128 & 1.127 & 1.130 & 1.137 & 1.121 & 1.079\\
   1.0 &  10 & 0.646 & 0.823 & 0.915 & 0.996 & 1.058 & 1.092 & 1.103 & 1.111 & 1.121 & 1.131 & 1.117 & 1.077\\
   1.0 &  25 & - & 0.416 & 0.709 & 0.871 & 0.977 & 1.040 & 1.069 & 1.087 & 1.107 & 1.123 & 1.112 & 1.073\\
   1.0 &  50 & - & - & - & 0.663 & 0.850 & 0.961 & 1.014 & 1.050 & 1.085 & 1.108 & 1.102 & 1.068\\
   1.0 & 100 & - & - & - & - & 0.595 & 0.811 & 0.910 & 0.976 & 1.042 & 1.080 & 1.085 & 1.057\\
   9.5 &   0 & 0.857 & 0.877 & 0.910 & 0.955 & 1.003 & 1.044 & 1.068 & 1.089 & 1.113 & 1.119 & 1.109 & 1.074\\
   9.5 &  10 & 0.532 & 0.683 & 0.791 & 0.882 & 0.955 & 1.013 & 1.047 & 1.075 & 1.104 & 1.113 & 1.105 & 1.072\\
   9.5 &  25 & - & 0.386 & 0.631 & 0.780 & 0.888 & 0.970 & 1.018 & 1.089 & 1.090 & 1.105 & 1.100 & 1.069\\
   9.5 &  50 & - & - & - & 0.610 & 0.784 & 0.904 & 0.970 & 1.021 & 1.068 & 1.090 & 1.091 & 1.063\\
   9.5 & 100 & - & - & - & - & 0.570 & 0.775 & 0.878 & 0.954 & 1.026 & 1.063 & 1.074 & 1.053\\

\enddata
\tablecomments{Radii of planets, in \rj.  Row 1 column headers are planet masses, in \mj, while row 2 is in \me.  The symbol ``-'' indicates that the planet mass is smaller than the given core mass.}
\label{t1000}
\end{deluxetable}

\begin{deluxetable}{cccccccccccccc}
\tabletypesize{\footnotesize}
%\rotate
\center
\tablecolumns{14}
\tablewidth{0pc}
\tablecaption{Giant Planet Radii at 4.5 Gyr}
\tablehead{
\colhead{Distance} & \colhead{Core} & \colhead{0.052} & \colhead{0.087} & \colhead{0.15} & \colhead{0.24} & \colhead{0.41} & \colhead{0.68} & \colhead{1.0} & \colhead{1.46} & \colhead{2.44} & \colhead{4.1} & \colhead{6.8}  & \colhead{11.3} \\ \colhead{(AU)} & \colhead{Mass} & \colhead{17} & \colhead{28} & \colhead{46} & \colhead{77} & \colhead{129} & \colhead{215} & \colhead{318} & \colhead{464} & \colhead{774} & \colhead{1292} & \colhead{2154}  & \colhead{3594} }
\startdata

  0.02 &   0 & * & 1.355 & 1.252 & 1.183 & 1.190 & 1.189 & 1.179 & 1.174 & 1.170 & 1.178 & 1.164 & 1.118\\
  0.02 &  10 & 0.726 & 0.934 & 1.019 & 1.072 & 1.123 & 1.148 & 1.153 & 1.157 & 1.160 & 1.172 & 1.160 & 1.116\\
  0.02 &  25 & - & 0.430 & 0.756 & 0.928 & 1.032 & 1.091 & 1.116 & 1.131 & 1.145 & 1.163 & 1.155 & 1.112\\
  0.02 &  50 & - & - & - & 0.695 & 0.891 & 1.004 & 1.056 & 1.091 & 1.121 & 1.148 & 1.144 & 1.106\\
  0.02 & 100 & - & - & - & - & 0.613 & 0.841 & 0.944 & 1.011 & 1.076 & 1.118 & 1.125 & 1.095\\
 0.045 &   0 & 1.103 & 1.065 & 1.038 & 1.049 & 1.086 & 1.105 & 1.107 & 1.108 & 1.113 & 1.118 & 1.099 & 1.053\\
 0.045 &  10 & 0.599 & 0.775 & 0.878 & 0.964 & 1.029 & 1.069 & 1.083 & 1.092 & 1.104 & 1.112 & 1.095 & 1.050\\
 0.045 &  25 & - & 0.403 & 0.686 & 0.846 & 0.952 & 1.019 & 1.050 & 1.069 & 1.090 & 1.104 & 1.090 & 1.047\\
 0.045 &  50 & - & - & - & 0.648 & 0.831 & 0.942 & 0.996 & 1.033 & 1.068 & 1.090 & 1.081 & 1.042\\
 0.045 & 100 & - & - & - & - & 0.587 & 0.798 & 0.896 & 0.961 & 1.026 & 1.062 & 1.063 & 1.032\\
   0.1 &   0 & 1.068 & 1.027 & 1.005 & 1.024 & 1.062 & 1.085 & 1.090 & 1.092 & 1.099 & 1.104 & 1.084 & 1.038\\
   0.1 &  10 & 0.592 & 0.755 & 0.858 & 0.942 & 1.008 & 1.051 & 1.067 & 1.077 & 1.090 & 1.098 & 1.080 & 1.036\\
   0.1 &  25 & - & 0.404 & 0.675 & 0.829 & 0.934 & 1.002 & 1.034 & 1.054 & 1.077 & 1.090 & 1.075 & 1.033\\
   0.1 &  50 & - & - & - & 0.639 & 0.817 & 0.928 & 0.982 & 1.019 & 1.055 & 1.076 & 1.066 & 1.027\\
   0.1 & 100 & - & - & - & - & 0.582 & 0.788 & 0.884 & 0.949 & 1.014 & 1.049 & 1.049 & 1.018\\
   1.0 &   0 & 1.014 & 0.993 & 0.983 & 1.011 & 1.050 & 1.074 & 1.081 & 1.084 & 1.091 & 1.096 & 1.075 & 1.030\\
   1.0 &  10 & 0.576 & 0.738 & 0.845 & 0.931 & 0.997 & 1.041 & 1.058 & 1.068 & 1.082 & 1.090 & 1.072 & 1.028\\
   1.0 &  25 & - & 0.400 & 0.666 & 0.820 & 0.924 & 0.993 & 1.026 & 1.046 & 1.069 & 1.082 & 1.067 & 1.025\\
   1.0 &  50 & - & - & - & 0.633 & 0.810 & 0.920 & 0.974 & 1.011 & 1.048 & 1.068 & 1.058 & 1.020\\
   1.0 & 100 & - & - & - & - & 0.578 & 0.782 & 0.878 & 0.942 & 1.007 & 1.041 & 1.041 & 1.010\\
   9.5 &   0 & 0.798 & 0.827 & 0.866 & 0.913 & 0.957 & 0.994 & 1.019 & 1.037 & 1.056 & 1.062 & 1.055 & 1.023\\
   9.5 &  10 & 0.508 & 0.653 & 0.759 & 0.844 & 0.911 & 0.966 & 0.999 & 1.024 & 1.048 & 1.057 & 1.052 & 1.021\\
   9.5 &  25 & - & 0.378 & 0.611 & 0.750 & 0.849 & 0.926 & 0.972 & 1.037 & 1.035 & 1.050 & 1.047 & 1.018\\
   9.5 &  50 & - & - & - & 0.594 & 0.754 & 0.865 & 0.926 & 0.973 & 1.015 & 1.037 & 1.039 & 1.013\\
   9.5 & 100 & - & - & - & - & 0.558 & 0.746 & 0.842 & 0.911 & 0.976 & 1.012 & 1.023 & 1.004\\

\enddata
\tablecomments{Radii of planets, in \rj.  Row 1 column headers are planet masses, in \mj, while row 2 is in \me.  The symbol ``-'' indicates that the planet mass is smaller than the given core mass  Given the albedos we calculate (see \S4.3) the approximate $T_{\rm{eq}}$ value at each distance is, 1960 K (0.02 AU), 1300 K (0.045 AU), 875 K (0.1 AU), 260 K (1 AU), and 78 K (9.5 AU). The symbol ``-'' indicates that the planet mass is smaller than the given core mass.}
\label{t4500}
\end{deluxetable}

\clearpage

\end{document}